\documentclass[aps,prb,twocolumn,footinbib,showpacs,superscriptaddress,preprintnumbers,amsmath,amssymb]{revtex4-1}

\usepackage{graphicx}
\usepackage{subfigure}
\usepackage{dcolumn}
\usepackage{bm}
\usepackage{verbatim} 
\usepackage{color}

\begin{document}

\title{Short time dynamics of molecular junctions after projective measurement}

\author{Gaomin Tang}
\affiliation{Department of Physics and the Center of Theoretical and Computational Physics, The University of Hong Kong, Pokfulam Road, Hong Kong, China}
\author{Yanxia Xing}
\affiliation{Beijing Key Laboratory of Nanophotonics and Ultrafine Optoelectronic Systems, School of Physics, Beijing Institute of Technology, Beijing 100081, China}
\author{Jian Wang}
 \email{jianwang@hku.hk}
\affiliation{Department of Physics and the Center of Theoretical and Computational Physics, The University of Hong Kong, Pokfulam Road, Hong Kong, China}

\date{\today}

\begin{abstract}
In this work, we study the short time dynamics of a molecular junction described by Anderson-Holstein model using full-counting statistics after projective measurement. The coupling between the central quantum dot (QD) and two leads was turned on at remote past and the system is evolved to steady state at time $t=0$, when we perform the projective measurement in one of the lead. Generating function for the charge transfer is expressed as a Fredholm determinant in terms of Keldysh nonequilibrium Green's function in the time domain. It is found that the current is not constant at short times indicating that the measurement does perturb the system. We numerically compare the current behaviors after the projective measurement with those in the transient regime where the subsystems are connected at $t=0$. The universal scaling for high-order cumulants is observed for the case with zero QD occupation due to the unidirectional transport at short times. The influences of electron-phonon interaction on short time dynamics of electric current, shot noise and differential conductance are analyzed.
\end{abstract}

\maketitle

\section{Introduction}
Quantum transport systems which are driven out of equilibrium due to external fields are stochastic in nature \cite{Blanter,Kampen}. Just as what is pointed out in the seminal paper by R. Landauer "The noise is the signal" \cite{Landauer}, cross current correlation can be used to determine whether the quasi-particle is fermionic or bosonic and one can get the effective charge of quasi-particles from the shot noise in fractional quantum Hall effect \cite{Saminadayar}. Full-counting statistics (FCS) in electronic transport which was initially formulated by Levitov and Lesovik can give us a full scenery of probability distribution of transferred charges besides the current and shot noise \cite{Levitov1,Levitov2,Levitov3, Levitov4, Klich1, Nazarov, RMP, Flindt1, wavepacket, exp1, Flindt2, Flindt3, Flindt4, gm1,gm2, yeyati_FCS, yeyati_FCS1, gm3, gm4, Ruben, Blatter3, negative_FCS1, negative_FCS2, gm6}. Generating function (GF), from which one can get high-order cumulants by taking derivatives with respect to the counting field, is the key in studying FCS and has various applications. Entanglement entropy is difficult to be measured experimentally, it was proposed that a series of the charge cumulants which are measurable can be used to approach it \cite{entangle1,entangle2,entangle3,entangle4,entangle5}. The dynamical Lee-Yang zeros of GF of an observable in open quantum systems can be accessed using high-order cumulants \cite{LYzeros1, LYzeros2, LYzeros3, LYzeros4}. The fluctuation theorem of GF can reveal the symmetry of a thermodynamic network \cite{RMP, Gaspard1, Hanggi, Matteo1, gm5}. Efficiency statistics of a thermoelectric engine can be calculated from the GF via the large deviation principle \cite{gm5, EF1, EF2, EF3, time_COP}. GF of spin transfer torque has also been used to calculate the magnetization switching probability \cite{switching}. Negative quasi-probability distributions is studied in FCS due to an interference effect \cite{negative_FCS1, negative_FCS2}.

	Besides FCS in steady states which has been studied extensively, FCS in transient regime attracts attentions recently \cite{gm1,gm2, yeyati_FCS, yeyati_FCS1, gm3, gm4, Ruben}. In the transient regime, the sub-systems are connected at time $t=0$ and then the connected system evolves towards a steady state. This is different from Cini's approach \cite{Cini, Wingreen, Maciejko} in which the system is well connected and in equilibrium and the bias is applied suddenly at $t=0$. Universal scaling behaviors with respect to relative amplitudes of the higher order particle or energy cumulants are found at short times for an initially unidirectional process \cite{yeyati_FCS, yeyati_FCS1, gm3, gm4}. FCS in the transient regime has also been used to determine the nonequilibrium population of the Andreev bound states in the quantum quench dynamics in Josephson junction \cite{Ruben}. Transient behaviors in cold atoms systems have also been investigated experimentally \cite{Atom1, Atom2}.

	For the projective measurement regime discussed in this work, the system was connected at remote past so that it reaches nonequilibrium steady state at $t=0$ after which we do quantum measurement \cite{RMP}. Discussions on quantum measurement in electronic transport systems involve both the von Neumann projecton postulate and detector's backaction on the system \cite{Blatter1, Blatter2}. In Refs.~[\onlinecite{Blatter1}] and [\onlinecite{Blatter2}], the quantum point contact (QPC) detector is capacitively coupled to the central scattering region and the current through QPC detector serves as a readout for the charge in the scattering region. When the coupling strength $t_c$ between the detector and scattering region is larger than $h\Gamma$ which is related to the QPC tunneling time scale $1/\Gamma$, strong backaction of the detector on the system leads to the strong projective measurement. The weak measurement regime goes to the case with $t_c\ll h\Gamma$. However, in the two-time measurement scheme in Ref.~[\onlinecite{RMP}], the strong (projective) measurement is performed in the electrode at two times, so that the number of electrons transferred during this period is counted. FCS of projective measurement in phonon transport system has been studied using this two-time measurement scheme \cite{RMP, JS1,JS2,JS3} and the lacuna should be filled in electron transport.

	In this work, we apply the Keldysh NEGF technique to investigate the short time behavior of electronic transport of a molecular junction with electron-phonon interaction after quantum projective measurement. Dressed tunneling approximation (DTA) is used in dealing with the the strong electron-phonon interaction \cite{BDong, DTA, BDong_SC}. GF is expressed in terms of a Fredholm determinant in the time domain.
	An approximate current expression is obtained from GF by expanding the determinant to first order with respect to the self-energy. This current approximation agrees quite well with the exact numerical derivative one. For the empty dot occupation case, transient regime where the sub-systems are suddenly connected has a a very good agreement with the dynamics after projective measurement, and the universal scaling for high-order cumulants is observed as well after projective measurement. Short time dynamics of current, shot noise and differential conductance after projective measurement are studied in the numerical section. The polaron effect on the current and shot noise will be discussed.
The differential conductance undergoes a sequence of steps and oscillations are observed at times $t\sim 2n\pi/\omega_0$ at different voltage threshold.
	 The rest of the article is organized as follows. In Sec. II, we present the Anderson-Holstein model of a molecular junction and GF expressed as a Fredholm determinant. Numerical results indicating the short time dynamics are shown in Sec. III. Finally, concluding remarks are made in Sec. IV.

\section{Model and theoretical formalism}

\subsection*{A. Model}
Molecular electronic devices, wherein the the electron-phonon interactions become pronounced, have been the focus of many investigations, both experimentally and theoretically \cite{molecule1, molecule2,FC_blockade1, FC_blockade2, molecule3, Schmidt, molecule4}. A variety of intriguing properties, such as negative differential conductance \cite{molecule2}, phonon-assisted current steps \cite{FC_blockade1, FC_blockade2}, Franck-Condon blockade \cite{FC_blockade1, FC_blockade2, molecule3}, and sign change in the shot noise correction \cite{Schmidt} have been found due to the interplay of electronic and vibrational degrees of freedom. Theoretically, these phenomena could be understood using a quantum dot (QD) described by the Anderson-Holstein model \cite{Egger, Holstein, Mahan} coupled to two electrodes.
	Considering only the lowest electronic orbital, the single-molecule can be simplified as a single electronic level of a QD being coupled to localized vibrational mode. The QD then is coupled to the left and right lead so that the system is driven to a nonequilibrium state with an external bias applied. The corresponding Hamiltonian reads as
\begin{equation}
H = H_{S} + H_L + H_R + H_T ,
\end{equation}
with the QD Hamiltonian (in natural units, $\hbar = k_B =e = m_e =1$)
\begin{equation}
H_{S} = \epsilon_0 d^\dag d + \omega_0 a^\dag a + t_{ep}(a^\dag +a) d^\dag d,
\end{equation}
where $\epsilon_0$ is the bare electronic energy level, and $\omega_0$ is the frequency of the localized vibron. $d^\dag$ ($a^\dag$) denotes the electron (phonon) creation operator in the QD. The localized vibron modulates the QD with the electron-phonon coupling constant $t_{ep}$. The Hamiltonian of the $\alpha$-lead is given by
\begin{equation}
H_\alpha = \sum_k \epsilon_{k\alpha} c_{k\alpha}^\dag c_{k\alpha} ,
\end{equation}
where the indices $k\alpha$ are used to label the different states in the left or right lead. $H_T$ describes the coupling between the dot and the leads with the tunneling amplitudes $t_{k\alpha}$,
\begin{equation}
H_T = H_{LS} + H_{RS}
= \sum_{k\alpha} (t_{k\alpha} c_{k\alpha}^\dag d + t_{k\alpha}^* d^\dag c_{k\alpha} ) .
\end{equation}
The tunneling rate (linewidth function) between QD and lead $\alpha$ is assumed to be Lorentzian and has the expression,
\begin{equation} \label{linewidth}
{\bf \Gamma}_\alpha(\omega) ={\rm Im} \sum_k \frac{|t_{k\alpha}|^2}{\omega-\epsilon_{k\alpha}-i0^+} = \frac{\Gamma_\alpha W^2}{\omega^2+W^2} ,
\end{equation}
with the linewidth amplitude $\Gamma_\alpha$ and bandwidth $W$, and one can denote $\Gamma = \Gamma_L+\Gamma_R$.
Applying the Lang-Firsov unitary transformation given by \cite{Lang-Firsov}
\begin{equation}
 \bar{H} = S H S^\dag, \ \ S=e^{g d^\dag d (a^\dag -a)},\ \ g = \frac{t_{ep}}{\omega_0},
\end{equation}
one can eliminate the electron-vibron interaction term and get the transformed Hamiltonian
\begin{equation}
\bar{H}_{S} = \bar{\epsilon} d^\dag d + \omega_0 a^\dag a ,
\end{equation}
with the effective bare QD electronic level $\bar{\epsilon}=\epsilon_0 - g^2\omega_0$. The tunneling Hamiltonian is then transformed to be
\begin{equation}
\bar{H}_T = \sum_{k\alpha } (t_{k\alpha} c_{k\alpha}^\dag X d + t_{k\alpha}^* d^\dag X^\dag c_{k\alpha})
\end{equation}
with the phonon cloud operator $X=\exp[g(a-a^\dag)]$, while Hamiltonians of isolated leads are not changed.

	Once the voltage bias is applied across the molecular junction, the system is under a non-equilibrium state and the particles transfer from one lead to the other. FCS can be used to characterize the probability distribution of transferred number of particles $\Delta n$ between an initial time $t=0$ and a later time $t$. The continuous GF $Z(\lambda, t)$ with the counting field $\lambda$ is defined as the Fourier transform of discrete probability distribution $P(\Delta n,t)$ and has the form,
\begin{equation} \label{Z}
Z(\lambda,t)=\sum_{\Delta n}P(\Delta n,t)e^{i\lambda\Delta n}.
\end{equation}
The $k$th charge cumulant $\langle\langle(\Delta n)^k\rangle\rangle$ can be calculated by taking the $k$th derivative of the cumulant generating function (CGF) which is the logarithm of GF with respect to $\lambda$ at $\lambda=0$:
\begin{equation} \label{kth}
C_k(t) \equiv \langle\langle(\Delta n)^k\rangle\rangle=\frac{\partial^k \ln Z(\lambda,t)}{\partial(i\lambda)^k}\bigg|_{\lambda=0} .
\end{equation}
The current cumulants which are defined as,
\begin{equation} \label{kth1}
\langle\langle I^k \rangle\rangle =\frac{\partial C_k(t)}{\partial t},
\end{equation}
tend to the steady state current cumulants in the long time limit $t\rightarrow \infty$. The second cumulant could be expressed as $C_2 (t) = \int_0^t dt_1 \int_0^t dt_2 \langle \delta I(t_1)\delta I(t_2) \rangle$, so that the second current cumulant expressed in a symmetry form is
\begin{equation}
\langle\langle I^2 \rangle\rangle = \frac{1}{2}\int_0^t dt_1 \left[ \langle \delta I(t_1)\delta I(t) \rangle + \langle \delta I(t) \delta I(t_1) \rangle \right] ,
\end{equation}
with $\delta I(t) = I(t)-\langle I(t)\rangle$.
One should note that the second current cumulant $\langle\langle I^2 \rangle\rangle$ is not an average of a squared quantity.  $\langle\langle I^2 \rangle\rangle$ is the zero frequency shot noise in the long time limit \cite{Blanter}.

\subsection*{B. Projective measurement and generating function}
We count the number of transferred electrons in the left lead, and the electrons flowing from the left lead to the QD is defined as the positive direction of the current. The current operator is given by
\begin{equation}
\hat{I}_L(t)=-d_t N^{(h)}_L(t) ,
\end{equation}
with the electron number operator $N_L^{(h)}(t)=\sum_{k}c^\dag_{kL}(t)c_{kL}(t)$ in the Heisenberg picture and $d_t$ being the total differential with respect to time. $N_L^{(h)}(t)$ is related to the number operator in the Schr\"{o}dinger picture $N_L(0)$ by,
\begin{equation}
N^{(h)}_L(t)=U(0,t)N_L(0)U(t,0),
\end{equation}
where the evolution operator is
\begin{equation}
U(t,t')=\mathbb{T}_C\exp\left\{-\frac{i}{\hbar}\int_{t'}^tH(t_1)dt_1\right\}, \quad (t>t'),
\end{equation}
with the time-ordering operator $\mathbb{T}_C$.
	The system starts at $t=-\infty$ with the leads and QD disconnected. The couplings between the leads and QD are switched on from $t=-\infty$ and the system evolves to steady state up to time $t=0$. This is different from the transient regime studied before \cite{gm1,gm2,gm3,yeyati_FCS} in which the couplings between the leads and central QD are suddenly turned on at $t=0$ and then system evolves towards the stationary state. In the regime considered in this work, the system at $t=0$ is in steady state $|\Psi_0\rangle$ and has a complete set of eigenstates $|n_0\rangle$ corresponding to number operator $N_L(0)$, that is,
\begin{equation}
N_L(0)|n_0\rangle =n_0 |n_0\rangle, \quad \  P_0=|n_0\rangle\langle n_0| .
\end{equation}
The system will be projected to state $P_0|\Psi_0\rangle$ after the first measurement at $t=0$ in the left lead. Second measurement with projective operator $P_t=|n_t\rangle\langle n_t|$ is performed at a later time $t$ on the evolved state $U(t,0) P_0 |\Psi_0\rangle$, so that the state at time $t$ is $|\Psi_t\rangle = P_t U(t,0) P_0|\Psi_0\rangle$.

\begin{figure}
  \includegraphics[width=3.50in]{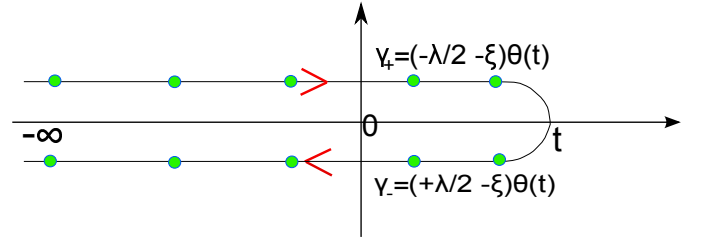}\\
  \caption{(Color online) Complex contour defined from time $-\infty$ to time $t$ and then back to time $-\infty$ in Keldysh space. }
  \label{fig1}
\end{figure}

	GF can be expressed over the Keldysh contour in this the two-time quantum measurement scheme as\cite{RMP, JS3, gm2, gm4},
\begin{equation}\label{Z1}
Z(\lambda,t) =\mathrm{Tr}\left\{ \rho'(0) U^\dag_{\lambda/2} (t,0) U_{-\lambda/2} (t,0) \right\},
\end{equation}
with the modified evolution operator ($\gamma = \pm \lambda/2$ depending on the branch of the contour, see Fig.~\ref{fig1}),
\begin{equation}\label{U}
U_\gamma(t,0) = \mathbb{T}_C \exp\left[ -\frac{i}{\hbar}\int_{0}^{t} H_\gamma(t') dt'\right] ,
\end{equation}
and the projected density matrix
\begin{equation}
\rho'(0)=\sum_{n_0}P_0\rho(0)P_0.
\end{equation}
The projected density matrix are used to measure the distribution of $N(0)$, the information of $\rho(0)$ is evolved from remote past and is unknown.
The modified Hamiltonian reads as $H_\gamma(t)=e^{i\gamma N_L(0)}\bar{H} e^{-i\gamma N_L(0)}$. Using the Baker-Hausdorff lemma, we obtain
\begin{align}
H_\gamma(t)=&\bar{H}_{S} + H_L +H_R \notag \\
&+ \sum_{k} \left[ e^{i\gamma} t_{kL}c_{kL}^\dag X d +t_{kR}c_{kR}^\dag X d + {\rm H.c.} \right] .
\end{align}
Here we should note that counting field $\gamma$ only enters the modified Hamiltonian through the coupling term between QD and the left lead in which the transferred electrons are counted.

	Projection operator $P_0$ could be written in the form as,
\begin{equation}
P_0 =\int_0^{2\pi}\frac{d\xi}{2\pi}e^{-i\xi(n_0-N_L(0))}  ,
\end{equation}
through Kronecker delta function, so that $\rho'(0)$ could be expressed in an integral form,
\begin{equation}  \label{rho'}
\rho'(0) =\int_0^{2\pi}\frac{d\xi}{2\pi} e^{i\xi N_L(0)}\rho(0)e^{-i\xi N_L(0)} .
\end{equation}
Plugging Eq.~\eqref{rho'} into Eq.~\eqref{Z1}, GF could be written as,
\begin{align}  \label{eq23}
Z(\lambda,t) =\int_0^{2\pi}\frac{d\xi}{2\pi}Z(\lambda,\xi,t) ,
\end{align}
with
\begin{equation}
Z(\lambda,\xi,t) =\mathrm{Tr}\left\{\rho(0)U_{\lambda /2-\xi}(0,t)U_{-\lambda /2-\xi}(t,0)\right\}.
\end{equation}
Since the coupling between QD and the leads is turned on at remote past $t=-\infty$, the density matrix $\rho(0)$ can be obtained by evolving the system from direct product state $\rho(-\infty)=\rho_L\otimes\rho_S\otimes\rho_R$ and expressed as,
\begin{equation}
\rho(0)=U(0,-\infty)\rho(-\infty)U(-\infty,0) .
\end{equation}
This enables us to rewrite Eq.~\eqref{eq23} as
\begin{align}
Z(\lambda,t)&= \int_0^{2\pi}\frac{d\xi}{2\pi}Z(\lambda,\xi,t) \notag  \\
&=\int_0^{2\pi}\frac{d\xi}{2\pi}{\rm Tr}\left\{\rho(-\infty)U_{\gamma_-}(-\infty,t)U_{\gamma_+}(t,-\infty)\right\} .
\end{align}
As shown in Fig.~(\ref{fig1}), the counting fields take values of \cite{gm2}
\begin{equation}
\gamma_+(t)=(-\lambda/2-\xi) \theta(t), \quad \gamma_-(t)=(\lambda/2-\xi) \theta(t),
\end{equation}
for the upper and lower branch of the Keldysh contour, respectively. Heaviside step function $\theta(t)$ is added due to the fact that the first measurement is performed at $t=0$.

	In the absence of electron-phonon interaction, GF is expressed as a Fredholm determinant in the time domain as \cite{RMP, gm2},
\begin{equation}  \label{GF}
Z(\lambda,\xi,t)=\det(G \widetilde{G}^{-1}).
\end{equation}
with
\begin{equation}  \label{Dyson}
\widetilde{G}^{-1}=G_0^{-1}-\widetilde{\Sigma}_L-\Sigma_R ,\quad
G^{-1}=G_0^{-1}-\Sigma_L-\Sigma_R  .
\end{equation}
$G_0(\tau,\tau')$ denotes the Green's function of the isolated QD, $\Sigma_\alpha$ is the self-energy due to the $\alpha$-lead, and the \textit{tilde} upon the self-energy indicates the inclusion of the counting field. The Green's functions and self-energies undergo the Keldysh structure as,
\begin{equation}
A=\begin{pmatrix}
A^{++} & A^{+-} \\ A^{-+} & A^{--}
\end{pmatrix} .
\end{equation}
Dyson equation defined on the Keldysh contour has the following relation (it also holds after Keldysh rotation which will be discussed later),
\begin{align}
G(t_1,t_2)=& G_0(t_1,t_2) \notag \\
&+\int_{-\infty}^t dt_3 \int_{-\infty}^t dt_4 G_0(t_1,t_3)\Sigma(t_3,t_4)G(t_4,t_2) ,
\end{align}
with $\Sigma(t_3,t_4)=\Sigma_L(t_3,t_4)+\Sigma_R(t_3,t_4)$.
Different components of left lead self-energy with counting field can be expressed by $\widetilde{\Sigma}_L^{ab}(t_1,t_2)=\exp[-i(\gamma_a-\gamma_b)]\Sigma_L^{ab}(t_1,t_2)$ with $a,b=+,-$ denoting different component index. Explicitly, when $-\infty<t_1<0$, $0<t_2<t$,
\begin{equation}
\widetilde{\Sigma}_L(t_1,t_2)
= e^{-i\xi} \begin{pmatrix}
e^{-i\lambda/2}\Sigma^{++}_L   &  e^{i\lambda/2} \Sigma^{+-}_L  \\
e^{-i\lambda/2}\Sigma^{-+}_L   &  e^{i\lambda/2}\Sigma^{--}_L
\end{pmatrix}_{(t_1,t_2)} ;
\end{equation}
and when $0<t_1<t$, $-\infty<t_2<0$, we can write $\widetilde{\Sigma}_L(t_1,t_2)$ as:
\begin{equation}
\widetilde{\Sigma}_L(t_1,t_2)
= e^{+i\xi} \begin{pmatrix}
e^{i\lambda/2}\Sigma^{++}_L   &  e^{i\lambda/2}\Sigma^{+-}_L  \\
e^{-i\lambda/2}\Sigma^{-+}_L   &  e^{-i\lambda/2}\Sigma^{--}_L
\end{pmatrix}_{(t_1,t_2)} ;
\end{equation}
and	when $0<t_1,t_2 <t$,
\begin{equation}
\widetilde{\Sigma}_L(t_1,t_2)
= \begin{pmatrix}
\Sigma^{++}_L   &  e^{i\lambda}\Sigma^{+-}_L  \\
e^{-i\lambda}\Sigma^{-+}_L   &   \Sigma^{--}_L
\end{pmatrix}_{(t_1,t_2)} ;
\end{equation}
and finally, when $-\infty<t_1, t_2<0$, $\widetilde{\Sigma}_L(t_1,t_2)=\Sigma_L(t_1,t_2)$.

	We now discuss the GF of the interacting case with electron-phonon interaction within dressed tunneling approximation (DTA) \cite{BDong, DTA, yeyati_FCS, BDong_SC, gm4}. Perturbative expansion is usually used when the electron-phonon interaction is weak \cite{perturb1,perturb2,perturb3} and it breaks down in dealing with strong electron-phonon interaction. Once the lifetime of the electronic state in the dot is much larger than that in the bridges between the leads and QD which is satisfied in the polaronic regime, we can apply DTA in which the leads' self-energies are dressed with the phonon cloud after decoupling the phonon cloud operator. DTA can eliminate the pathological features at low frequencies using the single particle approximation and at high frequencies using polaron tunneling approximation\cite{DTA}. The dressed self-energies under DTA are expressed as,
\begin{equation}
\Sigma_{d\alpha}^{ab} (t_1,t_2) = \Sigma_{\alpha}^{ab} (t_1,t_2) \Lambda^{ab} (t_1,t_2) .
\end{equation}
At zero-temperature, the lesser and greater components of phonon cloud operator $\Lambda(t_1,t_2)=\langle \mathbb{T}_C X^\dag(t_2)X(t_1)\rangle$ are given by \cite{Mahan},
\begin{equation} \label{phonon}
\Lambda^{+-}(t_1,t_2) = \left[\Lambda^{-+}(t_1,t_2)\right]^*
=\sum_{m\in \mathbb{N}} \alpha_{m} e^{i m\omega_0 (t_1-t_2)},
\end{equation}
with $\alpha_m = e^{-g^2} g^{2m}/{m!}$.
The rest components of $\Lambda$ could be obtained by the relations,
\begin{align}
&\Lambda^{++}(t_1,t_2)=\theta(t_1-t_2)\Lambda^{-+}(t_1,t_2)+\theta(t_2-t_1)\Lambda^{+-}(t_1,t_2), \notag \\
&\Lambda^{--}(t_1,t_2)=\theta(t_2-t_1)\Lambda^{-+}(t_1,t_2)+\theta(t_1-t_2)\Lambda^{+-}(t_1,t_2).
\end{align}
Self-energy $\Sigma_\alpha$ is replaced with $\Sigma_{d\alpha}$ in the Dyson equation with the expression,
\begin{equation} \label{dyson1}
G = G_0 + G_0 \Sigma_{d} G ,
\end{equation}
where $\Sigma_{d} = \Sigma_{dL}+ \Sigma_{dR}$. GF in the strong electron-phonon coupling under DTA is similar with Eq.~\eqref{GF}, so that
\begin{equation} \label{GF1}
Z(\lambda,t)=\int_0^{2\pi}\frac{d\xi}{2\pi}\det(G \widetilde{G}^{-1}),
\end{equation}
with
\begin{equation}
\widetilde{G}^{-1}=G_0^{-1}-\widetilde{\Sigma}_{dL}-\Sigma_{dR} .
\end{equation}
	
	One can also perform Keldysh rotation \cite{Keldysh, Ka2, gm2} to transform the Green's function and self-energies into upper triangular matrices in Keldysh space (Larkin-Ovchinnikov ones) with the relation
\begin{equation}
\begin{pmatrix}
A^r & A^k \\ 0 & A^a
\end{pmatrix} =
L\sigma_x\begin{pmatrix}
A^r & A^k \\ 0 & A^a
\end{pmatrix} L^{-1} ,
\end{equation}
where the Keldysh matrix is
\begin{equation}
L=\frac{1}{\sqrt{2}}
\begin{pmatrix}
1 & 1 \\ -1 & 1
\end{pmatrix} .
\end{equation}
The dressed retarded self-energy can be calculated through the relation $\Sigma_{d\alpha}^r(t_1,t_2)=\theta(t_1-t_2)\left[\Sigma_{d\alpha}^{+-}(t_1,t_2)-\Sigma_{d\alpha}^{-+}(t_1,t_2)\right]$, and the dressed Keldysh component of self-energy is $\Sigma_{d\alpha}^k = 2\Sigma_{d\alpha}^< +\Sigma_{d\alpha}^r-\Sigma_{d\alpha}^a$ with $\Sigma_{d\alpha}^<=-\Sigma_{d\alpha}^{+-}$.
Due to Keldysh rotation, the left lead self-energy with counting field reads \cite{gm2}
\begin{equation}
\widetilde{\Sigma}_{dL}(t_1,t_2)
= \Upsilon^*[\gamma(t_1)]
\begin{pmatrix}
    \Sigma^{r}_{dL}  &  \Sigma^{k}_{dL}   \\
     0            &  \Sigma^{a}_{dL}
\end{pmatrix}_{(t_1,t_2)}
\Upsilon[\gamma(t_2)] ,
\end{equation}
with $\Upsilon[\gamma(\tau)]=\exp(-i\xi)\exp(-i\lambda\sigma_x/2)$ if $\tau\geq 0$ and $\Upsilon[\gamma(\tau)]=1$ for $\tau <0$.
	The self-energy $\widetilde{\Sigma}_{dL}(\tau,\tau')$ in the presence of the counting field should be calculated separately at four different time regions. Explicitly, when $-\infty<t_1<0$, $0<t_2<t$, ($\Sigma^r_L=0$), we can write $\widetilde{\Sigma}_{dL}(t_1,t_2)$ as,
\begin{equation}
\widetilde{\Sigma}_{dL}(t_1,t_2)
= e^{-i\xi} \begin{pmatrix}
  -i\sin\frac{\lambda}{2}\Sigma^k_{dL}   & \cos\frac{\lambda}{2}\Sigma^k_{dL}  \\
  -i\sin\frac{\lambda}{2}\Sigma^a_{dL}   &   \cos\frac{\lambda}{2}\Sigma^a_{dL}
\end{pmatrix}_{(t_1,t_2)} ;
\end{equation}
and when $0<t_1<t$, $-\infty<t_2<0$, ($\Sigma^{a}_L=0$):
\begin{equation}
\widetilde{\Sigma}_{dL}(t_1,t_2)
= e^{+i\xi} \begin{pmatrix}
  \cos\frac{\lambda}{2}\Sigma^r_{dL}  & \cos\frac{\lambda}{2}\Sigma^k_{dL}   \\
 i\sin\frac{\lambda}{2}\Sigma^r_{dL} & i\sin\frac{\lambda}{2}\Sigma^k_{dL}
\end{pmatrix}_{(t_1,t_2)} ;
\end{equation}
and	when $0<t_1,t_2 <t$, $\widetilde{\Sigma}_{dL}(t_1,t_2)$ has the expression as \cite{entries}
\begin{equation}
\widetilde{\Sigma}_{dL}(t_1,t_2)= \exp(i\lambda\sigma_x/2)\Sigma_{dL}(t_1,t_2) \exp(-i\lambda\sigma_x/2).
\end{equation}
The Keldysh transformation facilitates us in calculating GF numerically since the Green's function and self-energies without counting field are upper-triangular in Keldysh space, the determinant $\det(G)$ could be calculated by directly multiplying its diagonal entries. In the numerical calculations, one can get the diagonal elements of $\widetilde{G}^{-1}$ first so that the inverse of all these diagonal elements constitute a diagonal matrix $\delta$ and $\det(G \widetilde{G}^{-1})=\det(\delta \widetilde{G}^{-1})$. The time slice discretization of Green's function and self-energies in time domain could be found in Refs. [\onlinecite{gm2}] and [\onlinecite{gm4}].

	Taking the derivative of the GF, Eq.~\eqref{GF1}, with respect to $i\lambda$ using Jacobi's formula and expanding the determinant to first order in the self-energy, we can get an approximate expression of the average number of transferred electrons,
\begin{align}
&\langle \Delta n_L(t) \rangle = \int_0^{2\pi} \frac{d\xi}{2\pi} \frac{\partial}{\partial(i\lambda)}\det(G \widetilde{G}^{-1})\bigg|_{\lambda=0}  \notag \\
&\approx \int_{0}^{t}d\tau\int_{0}^{\tau}d\tau'
[G^r(\tau,\tau')\Sigma_{dL}^<(\tau',\tau)+G^<(\tau,\tau')\Sigma_{dL}^a(\tau',\tau)] \notag   \\
&- \int_{0}^{t}d\tau\int_{0}^{\tau}d\tau'
[G^a(\tau',\tau)\Sigma_{dL}^<(\tau,\tau')+G^<(\tau',\tau)\Sigma_{dL}^r(\tau,\tau')] .
\end{align}
From $\langle \Delta n_L(t) \rangle=\int_0^t I_L(\tau) d\tau$, we obtain an approximate expression for the current in the left lead after projective measurement at time $t$,
\begin{align} \label{approx}
I_L(t)\approx\int_{0}^{t}d\tau [G^r(t,\tau)\Sigma_{dL}^<(\tau,t)+G^<(t,\tau)\Sigma_{dL}^a(\tau,t)]+{\rm H.c.}.
\end{align}
We can observe that the current expression is different from that of the steady state \cite{Haug} wherein the integral with respect to time is from $-\infty$ to $t$. Due to the absence of time translation invariance, the current in the left lead is not the same with the one in the right lead in short times. We will also numerically show that the current after projective measurement oscillates in the short time and evolves to the steady state value. This confirms the fact the system is perturbed after the first projective measurement. Similar behaviors of heat current have been studied previously in phonon transport.\cite{JS1,JS2}

\section{Numerical Results}
In this section, we show our numerical results at zero temperature. The linewidth amplitudes in the left and right lead are set equal with $\Gamma_L=\Gamma_R=0.5\Gamma$, and the bandwidth is set as $W=10\Gamma$ through all the calculations. The voltage bias $\Delta\mu=\mu_L-\mu_R$ is symmetrically applied to the left and right lead with $\mu_L=-\mu_R$. The Fredholm determinant is calculated in discretized time slice grid \cite{gm2,gm4}. Cumulants and current cumulants are measured in the left lead in default.

\begin{figure}
  \includegraphics[width=3.7in]{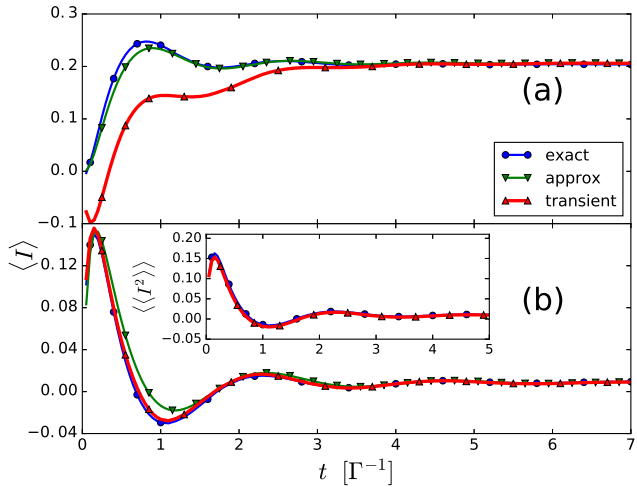}\\
  \caption{(Color online) Comparison of currents among the exact one after projective measurement (blue line), the approximation using Eq.~\eqref{approx} (green line), and transient regime (red line), in the absence of electron-phonon interaction. The energies are measured in the unit of $\Gamma$ and the voltage bias is $\Delta\mu=5\Gamma$. Two cases with different energy levels in QD are considered: $\bar{\epsilon}=-\Gamma$ [upper panel] and $\bar{\epsilon}=5\Gamma$ [lower panel]. $1/\Gamma$ is the unit of time. In the inset of lower panel, we also compare current noise between the projective measurement and transient regime for $\bar{\epsilon}=5\Gamma$. }
  \label{fig2}
\end{figure}

In Fig.~\ref{fig2}, we compare the currents among the exact one by numerical derivative with respect to $\lambda$ after projective measurement, the approximation using Eq.~\eqref{approx}, and transient regime for $\bar{\epsilon}=-\Gamma$ [upper panel] and $\bar{\epsilon}=5\Gamma$ [lower panel] in the absence of electron-phonon interaction. The voltage bias is $\Delta\mu=5\Gamma$ so that $\mu_L=2.5\Gamma$. For the transient regime, we shall have an initial dot occupation \cite{yeyati_FCS, gm3,gm4}, which is set related to the steady state lesser Green's function at equal times with $n_d = -iG^<(0,0)$. For the case in which the effective QD level is chosen between the leads chemical potentials with $\bar{\epsilon}=-\Gamma$, $n_d =0.8707$, and for $\bar{\epsilon}=5\Gamma$ which is above $\mu_L$, $n_d =0$. We can observe that currents oscillate at short times and the currents calculated from Eq.~\eqref{approx} have good approximations at both short times and long times in spite of some deviations near $t\approx \Gamma^{-1}$. The current and current noise (i.e., the second current cumulant, shown in the inset of lower panel) for $\bar{\epsilon}=5\Gamma$ calculated in the transient regime almost agree with the exact ones while for $\bar{\epsilon}=-\Gamma$ the currents agree poorly at short times. It suggests that the initial density matrix $\rho(0)$ is almost diagonal for $\bar{\epsilon}=5\Gamma$, implying that the coherence in the system is not important. We can conclude that the dynamics after projective measurement can be well described by the transient regime for the zero dot occupation case.

\begin{figure}
  \includegraphics[width=3.0in]{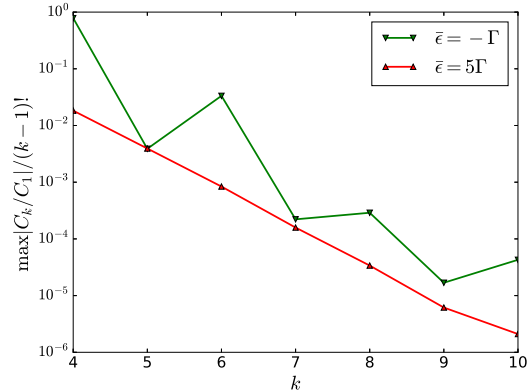}\\
  \caption{(Color online) The maximum amplitudes of the relative cumulants $C_k/C_1$ normalized with $(k-1)!$ in the logarithmic scale for $\bar{\epsilon}=-\Gamma$ (green line) and $\bar{\epsilon}=5\Gamma$ (red line).
  The linear slope corresponding to $\bar{\epsilon}=5\Gamma$ (zero dot occupation) indicates a universal scaling with ${\rm max}|C_k/C_1| \sim (k-1)! \ x^{-k}$ where $x$ in an unknown constant in this work. The universal scaling is broken for the case of $\bar{\epsilon}=-\Gamma$. }
  \label{fig3}
\end{figure}

We study the the universal scaling of high-order cumulants (from the $4$th to $10$th order) for both $\bar{\epsilon}=-\Gamma$ and $\bar{\epsilon}=5\Gamma$ in Fig.~\ref{fig3}. The cumulants are obtained by numerical derivatives with respect to $\lambda$ using Eqs.~\eqref{kth} and \eqref{GF1}. The maximum amplitudes of the relative cumulants $C_k/C_1$ normalized with $(k-1)!$ in the logarithmic scale are shown. The linear slope corresponding to $\bar{\epsilon}=5\Gamma$ ($n_d=0$) indicates a universal scaling of high-order cumulants with ${\rm max}|C_k/C_1| \sim (k-1)! \ x^{-k}$ where $x$ is an unknown constant ($x=\pi$ in the transient regime \cite{yeyati_FCS}). The universal scaling is broken for the case of $\bar{\epsilon}=-\Gamma$. A slight deviation from the linear slope of $10$th cumulant for $\bar{\epsilon}=5\Gamma$ may be due to numerical inaccuracy. The analytical explanation has been made in the transient regime that unidirectional transport is essential to have this linear slope of universal scaling \cite{yeyati_FCS, yeyati_FCS1}. This is also reported experimentally in measuring high-order cumulants in a steady state Coulomb blockade system \cite{exp1}. In the projective measurement discussed in this work, the short time behavior is unidirectional for $\bar{\epsilon}=5\Gamma$, in which it can be well described by the transient regime as shown in Fig.~\ref{fig2}, while bidirectional for $\bar{\epsilon}=-\Gamma$.

\begin{figure}
  \includegraphics[width=3.7in]{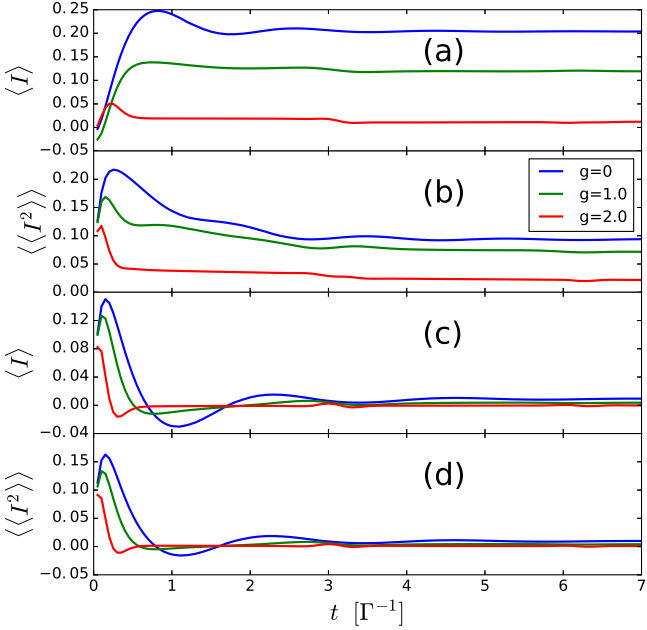}\\
  \caption{(Color online) Current and current noise for $\bar{\epsilon}=-\Gamma$ [panel (a) and panel (b)] and $\bar{\epsilon}=5\Gamma$ [panel (c) and panel (d)] by varying electron-phonon interaction constant $g=0$ (blue line), $g=1.0$ (green line), and $g=2.0$ (red line). }
  \label{fig4}
\end{figure}

In order to show the influences of electron-phonon interaction on the short time behaviors, we plot the current and current noise for $\bar{\epsilon}=-\Gamma$ [panel (a) and panel (b)] and $\bar{\epsilon}=5\Gamma$ [panel (c) and panel (d)] by varying electron-phonon interaction constant $g=0$, $g=1.0$, and $g=2.0$ in Fig.~\ref{fig4}. The currents at $t=0$ are finite which indicates the backation from the detector. The initial currents increase with a positive slope and then decrease and oscillate towards steady state values. With increased interaction constant $g$, the times when the maximum currents locate shift towards smaller times and eventually the initial slope becomes negative for large $g$.
For the case of $\bar{\epsilon}=5\Gamma$, the dips' positions of the current and noise shift towards smaller times with with increasing $g$, which is also reported in the transient regime \cite{yeyati_FCS}. One can also observe there are small steps at time $t \sim 2n\pi/\omega_0$ for $g=2$ due to the polaron dynamics and this becomes more pronounced for the case of $\bar{\epsilon}=5\Gamma$.

\begin{figure}
  \includegraphics[width=3.7in]{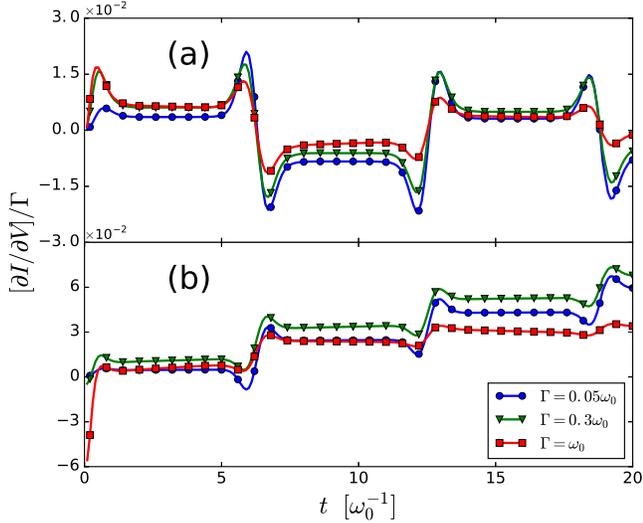}\\
  \caption{(Color online) The evolution of short time differential conductance $\partial I/\partial V$ (normalized by $\Gamma$) under different linewidth amplitude $\Gamma=0.05\omega_0$ (blue line), $\Gamma=0.25\omega_0$ (green line) and $\Gamma=\omega_0$ (red line). The voltage bias is set as $\Delta\mu=\omega_0$ in the upper panel, and $\Delta\mu=2\omega_0$ in the lower one. $\bar{\epsilon}=0$ and $g=2.0$.  }
  \label{fig5}
\end{figure}

The current behaviors at the inelastic thresholds with $\Delta\mu=n\omega_0$ have been investigated both in the steady state \cite{FC_blockade1, FC_blockade2}, and transient regime \cite{yeyati_FCS}. It is worth studying the short time differential conductance behaviors of the molecular junction after projective measurement. In Fig.~\ref{fig5}, we plot the evolution of short time differential conductance $\partial I/\partial V$ normalized by $\Gamma$ by varying linewidth amplitudes for $\Delta\mu=\omega_0$ [upper panel] and $\Delta\mu=2\omega_0$ [lower panel]. Electron-phonon interaction constant is chosen to be $g=2.0$. $\omega_0^{-1}$ is the unit of time. The effective QD level is set zero for a perfect transmitting junction so that the dot occupation is finite and the transport process at short times is not unidirectional. The differential conductance undergoes a sequence of up and down steps and only up steps for $\Delta\mu=2\omega_0$ at times $t\sim 2n\pi/\omega_0$. The oscillations of differential conductance at $t\sim 2n\pi/\omega_0$ are observed and these are absent in the transient regime \cite{yeyati_FCS}. For $\Delta\mu=2\omega_0$ case, the differential conductances start from negative values at very short times and can even oscillate to negative values during the evolution even though the the overall steps are upward. With the increasing linewidth amplitude $\Gamma$, the step structures together with the oscillations of the differential conductance are damped for both $\Delta\mu=\omega_0$ and $\Delta\mu=2\omega_0$.

\begin{figure}
  \includegraphics[width=3.7in]{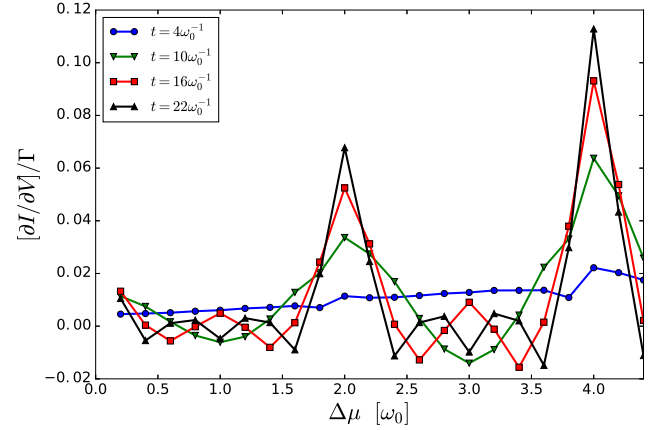}\\
  \caption{(Color online) Differential conductances $\partial I/\partial V$ versus applied voltage bias at different times. Since the differential conductance variation period is $2\pi/\omega_0$, $t=4\omega_0^{-1}$, $10\omega_0^{-1}$, $16\omega_0^{-1}$, and $22\omega_0^{-1}$ are located in the first, second, third, and fourth period, respectively. The bias voltage ranges from $0.2\omega_0^{-1}$ to $4.4\omega_0^{-1}$ with step size $0.2\omega_0^{-1}$. The discontinuity of each line is due to the large voltage step size. }
  \label{fig6}
\end{figure}

In Fig.~\ref{fig6}, we plot differential conductances $\partial I/\partial V$ versus applied voltage bias at different times. Since the conductance variation period is $2\pi/\omega_0$, $t=4\omega_0^{-1}$, $10\omega_0^{-1}$, $16\omega_0^{-1}$, and $22\omega_0^{-1}$ are located in the first, second, third, and fourth period, respectively. We can observe that the sign of the differential conductance plateau alternates and the amplitude decreases with period when the voltage bias is odd times of polaron frequency. Differential conductances are always positive in each plateau and the amplitude increases with period once the voltage bias is even times of polaron frequency.

\section{Conclusion}
We study the short time dynamics of a molecular junction described by Anderson-Holstein model using full-counting statistics after projective measurement, and obtained the GF expressed as a Fredholm determinant in the framework of NEGF by using DTA to deal with electron-phonon coupling. We perform the projective measurement in the left lead at time $t=0$ when the system has been in a nonequilibrium steady state. We obtain an approximate current expression from GF by expanding the determinant to first order with respect to the self-energy. This current approximation agrees quite well with the exact one which is obtained by taking the numerical derivative with respect to counting field $\lambda$. The comparison between the measurement and transient regime shows a very good agreement for both current and noise in the empty dot occupation case.
The universal scaling for high-order cumulants is observed for the case with zero QD occupation due to the unidirectional transport at short times, while the universal scaling is broken for the case with nonzero dot occupation.
Short time dynamics of electronic current, noise and differential conductance are analyzed in the presence of electron-phonon interaction. The currents at $t=0$ are finite which indicates the backation effect due to quantum measurement. The times when the maximum currents locate decrease with increasing electron-phonon interaction constant. The occurrences of small steps at time $t \sim 2n\pi/\omega_0$ for $g=2.0$ is due to the polaron dynamics and this becomes more pronounced for the case of zero dot occupation.
The differential conductance undergoes a sequence of steps and oscillations are observed at times $t\sim 2n\pi/\omega_0$ for both $\Delta\mu=\omega_0$ and $\Delta\mu=2\omega_0$. With increased linewidth amplitude, the step structures together with the oscillations of the differential conductances are progressively damped.

\begin{acknowledgements}
G. Tang and J. Wang were supported by NSF-China (Grant No. 11374246), the General Research Fund (Grant No. 17311116) and the University Grant Council (Contract No. AoE/P-04/08) of the Government of HKSAR.  Y. Xing was supported by NNSF project of China (Grant No. 11674024). We thank R. Seoane Souto for useful discussions.
\end{acknowledgements}


\begin{thebibliography}{99}
\bibitem{entries} The four entries of $\widetilde{\Sigma}_{dL}(t_1,t_2)$ with $0<t_1,t_2<1$ are respectively as,
\begin{align*}
\widetilde{\Sigma}_{dL}^{11}&= \cos^2\frac{\lambda}{2}\Sigma^r_{dL}-\frac{i}{2}\sin\lambda\Sigma^k_{dL}+\sin^2\frac{\lambda}{2}\Sigma^a_{dL} ,   \notag \\
\widetilde{\Sigma}_{dL}^{12}&= -\frac{i}{2}\sin\lambda\Sigma^r_{dL}+\cos^2\frac{\lambda}{2}\Sigma^k_{dL}+\frac{i}{2}\sin\lambda\Sigma^a_{dL}  ,  \notag \\
\widetilde{\Sigma}_{dL}^{21}&= +\frac{i}{2}\sin\lambda\Sigma^r_{dL}+\sin^2\frac{\lambda}{2}\Sigma^k_{dL}-\frac{i}{2}\sin\lambda\Sigma^a_{dL}  ,  \notag \\
\widetilde{\Sigma}_{dL}^{22}&= \sin^2\frac{\lambda}{2}\Sigma^r_{dL}+\frac{i}{2}\sin\lambda\Sigma^k_{dL}+\cos^2\frac{\lambda}{2}\Sigma^a_{dL} .
\end{align*}
\end{thebibliography}
\end{document}